\documentclass[epj]{svjour}
\usepackage{amssymb,graphicx,graphics}

\newcommand{\lao}{LaOFeAs}
\newcommand{\tn}{$T_{\rm{N}}$}
\newcommand{\ts}{$T_{\rm{S}}$}

\begin{document}
\title{Synthesis and physical properties of $\rm\bf LaO_{1-x}F_xFeAs$}
\author{Agnieszka Kondrat\inst{1}\thanks{These people contributed equally to this work.}, Jorge Enrique Hamann-Borrero\inst{1}$^{\rm a}$ \and  Norman Leps\inst{1}$^{\rm a}$ \and  Martin Kosmala\inst{2} \and  Olaf Schumann\inst{2} \and Anke K\"{o}hler\inst{1} \and Jochen Werner\inst{1} \and  G\"{u}nter Behr\inst{1} \and  Markus Braden\inst{2} \and  R\"{u}diger Klingeler\inst{1} \and  Bernd B\"uchner\inst{1} \and   Christian Hess\inst{1}\thanks{\email{c.hess@ifw-dresden.de}}}

%
%
\institute{Leibniz-Institute for Solid State and Materials Research, IFW-Dresden, 01171 Dresden, Germany \and II. Physikalisches Institut, Universit\"{a}t zu K\"{o}ln, 50937 K\"{o}ln, Germany}
\date{Received: date / Revised version: date}
%
\abstract{We have prepared the newly discovered Fe-based superconducting material $\rm LaO_{1-x}F_xFeAs$ ($0\leq x\leq 0.2$) in polycrystalline form and have investigated the samples by means of structural, thermodynamic and transport measurements. Our investigations reveal a non superconducting phase at $0\leq x\lesssim0.04$ which for $x=0$ is characterized by a structural transition towards an orthorhombic distortion at $T_s\approx 160$~K  and antiferromagnetic spin order at $T_N\approx138$~K. Both transitions lead to strong anomalies in various transport properties as well as in magnetization and in specific heat. Remarkably, the transition temperatures are only weakly doping dependent up $x\lesssim 0.04$. However, the transitions are abruptly suppressed at $x\geq0.05$ in favour of a superconducting phase with a critical temperature $T_c\gtrsim 20$~K. Upon further increasing the F-doping $T_c$ increases up to a maximum of $T_c=26.8$~K  at $x=0.1$ which is followed by a decrease down to $T_c\approx10$~K at $x\geq0.15$. 
\PACS{
      {74.70.-b}{Superconducting materials}   \and
      {74.25.Bt}{Thermodynamic properties} \and
      {74.25.Fy}{Transport properties (electric and thermal conductivity, thermoelectric effects, etc.)}
    } 
} 
\maketitle

\section{Introduction}
The ongoing search for new superconductors has recently yielded a new family of Fe-based compounds composed of alternating $\rm LaO_{1-x}F_x$ and $\rm FeAs$ layers with
transition temperatures $T_\mathrm{c}$ up to 28~K \cite{Kamihara2008}. Rapidly after the discovery, $T_\mathrm{c}$ has been raised to above 50~K \cite{Chen2008a,Cheng2008,Liu2008a,Ren2008b,Ren2008c} by replacing
La with other Rare Earths, and thus the first non-copper-oxide superconductor with $T_\mathrm{c}$ exceeding 50~K has emerged. 
Both, theoretical treatments and experimental findings indicate unconventional multiband superconductivity in these materials. Ab-initio calculations of the electron-phonon coupling are incompatible with a conventional electron-phonon pairing mechanism \cite{Boeri2008,Haule2008a,Mazin2008} which is consistent with experimental findings \cite{Drechsler2008}. The unusual nature of superconductivity is experimentally also evident from the absence of a Hebel-Slichter peak in NMR experiments \cite{Grafe2008} and high magnetic field experiments \cite{Fuchs2008a,Hunte2008}.
Various mechanism of superconductivity have been discussed and different pairing symmetries of the superconducting ground state including spin-triplet p-wave pairing have been proposed \cite{Mazin2008,Dai2008,Kuroki2008,Han2008,Lee2008}.
Intriguingly, there is evidence for a close interplay between superconductivity and magnetism as it is well established for other unconventional superconductors. A commensurate spin-density wave (SDW) and an orthorhombic distortion have been observed below $\sim150$~K in the undoped parent compound \cite{Cruz2008,Klauss2008} which appear to be fully suppressed once superconductivity emerges upon doping \cite{Luetkens2008,Luetkens2009a}. Note, that the SDW ground state has also been established for isostructural Rare Earth (R) based $\rm RO_{1-x}F_xFeAs$ \cite{Drew2009}, and even in the parent material of other iron-pnictide superconductors such as $\rm Ba_{1-x}K_xFe_2As_2$ which exhibit a different crystal structure but similar $\rm Fe_2As_2$ layers \cite{Rotter2008a,Rotter2008b,Rotter2008}. While the superconducting state of $\rm LaO_{1-x}F_xFeAs$ has been shown to exhibit no traces of static magnetic order \cite{Luetkens2008,Luetkens2009a}, the coexistence of static magnetic order and superconductivity has been reported \cite{Drew2009,Drew2008,Goko2008,Park2008}.

In this paper, we investigate the impact of the structural and magnetic transitions of undoped LaOFeAs on bulk physical properties and study the evolution as a function of doping. In particular we present investigations of $\rm LaO_{1-x}F_xFeAs$ by means of structural, thermodynamic and transport measurements. Our investigations reveal a non superconducting phase at $0\leq x\lesssim0.04$ where the physical properties are dominated by a two successive transitions. At higher temperatures ($T_s\approx 160$~K at $x=0$), there is a transition from tetragonal to orthorhombic structure which is accompanied by distinct drops in the magnetic susceptibility $\chi$ and the electrical resistivity $\rho$. At slightly lower temperature ($T_N\approx138$~K at $x=0$), a magnetic transition occurs towards long range spin density wave (SDW) antiferromagnetic order, which leads to an inflection point in the resistivity \cite{Klauss2008}. Remarkably these transition temperatures are only weakly doping dependent up to $x\lesssim 0.04$. However, the transitions are abruptly suppressed at $x\geq0.05$ in favour of a superconducting phase with a critical temperature $T_c\gtrsim 20$~K, i.e. close to the boundary in the phase diagram where superconductivity emerges. Upon increasing the F-doping, $T_c$ increases up to a maximum of $T_c=26.8$~K at $x=0.1$ which is followed by a decrease with $T_c\approx10$~K at $x\geq 0.15$. Intriguingly, for lower doping levels $0.05\leq x\leq0.075$ of these superconducting samples the temperature dependence of the resistivity at lower temperatures is reminiscent of that of the non-superconducting doping levels, i.e. $\rho(T)$ shows a weakly insulating behavior at $T\lesssim 80$~K. At higher doping levels $x\geq 0.1$, a quadratic temperature dependence evolves in the normal state just above $T_c$ with increasing doping.

\section{Experimental}
\subsection{Sample preparation and characterization}

\begin{figure}
\includegraphics [width=\columnwidth,clip] {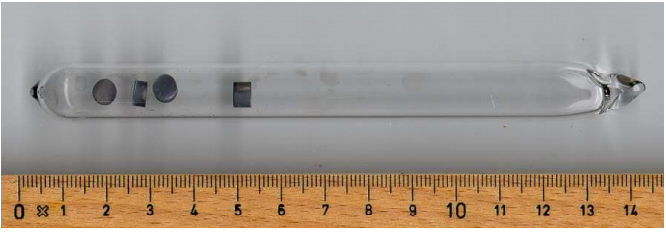}
\caption{Synthesis of $\rm LaO_{1-x}F_{x}FeAs$ polycrystalline samples in a silica ampoule.}
\label{fig:ampule}
\end{figure}
Polycrystalline samples of $\rm LaO_{1-x}F_xFeAs$ ($0\leq x\leq 0.2$) were prepared from pure components using a two-step solid state reaction method, similar to that described by Zhu et al. \cite{Zhu2008a}.  In the first step, Fe powder (Alfa Aesar, Puratronic, 99.998\%) and powdered As lump (Alfa Aesar, 99.999\%)  were milled, mixed and pressed into pellets and annealed at 500$^\circ$C for 2 h and at 700$^\circ$C for 10 h in an evacuated silica tube. In the second step, the FeAs pellets were milled and mixed with lanthanum powder (Goodfellow, 99.9\%, 60 mesh), annealed $\rm La_2O_3$ powder (Chempur, 99.99\%), and anhydrous $\rm LaF_3$ powder (Alfa Aesar, 99.99\%) and pressed into pellets under a pressure of 1 GPa. All preparation steps were carried out under argon atmosphere. The samples were then heated in an evacuated silica tube at 940$^\circ$C for 8 h and at 1150$^\circ$C for 40 h. Due to the formation of a liquid phase at 1007$^\circ$C, determined by DTA/TG analysis, the 940$^\circ$C annealing step was extended compared to Zhu et al. \cite{Zhu2008a} in order to improve the homogeneity.

Since the oxidation state of the system depends strongly on the oxygen stoichiometry, the oxygen content of the starting metals (La and Fe) was measured \cite{Souptel2007} and taken into account in the weighting procedure of the components. The amount of oxygen in the powders was found to be 2.8~at\% for Fe and 3.3~at\% for La, respectively. Note that these values differ from lot to lot of the same supplier.
A wrapping of the samples in Ta-foil during the annealing process as is described in the literature was not used because this induces an extended incorporation of As into the Ta-foil up to 4 wt\% which leads to As deficiency in the samples and, hence, changed physical properties of $\rm LaO_{1-x}F_xFeAs$ \cite{Fuchs2008a}. However, an elevated As vapour-pressure was observed during annealing which mainly depends on the three factors temperature, ampoule volume and sample mass, which allows to optimize the process. 

The appearance of the quartz ampoules after annealing (see Figure~\ref{fig:ampule} for a representative example) indicates no significant reaction of the pellets with the quartz wall, which confirms that annealing without Ta foil is possible. Only in the case of the high fluorine doped samples the quartz was slightly attacked.

The annealed pellets were ground and polished and the local composition of the resulting samples was investigated by wavelength-dispersive X-ray spectroscopy (WDX) in a scanning electron microscope.
Moreover, room temperature Mo $K_\alpha$ powder x-ray diffraction (XRD) in transmission geometry was performed on the samples in order to check the phase purity and to obtain structural information via  Rietveld refinement. The data was collected over a $2\theta$ range of $4^\circ$ to $55^\circ$ with a step size of $0.01^\circ$. The refined parameters include zero point offset, lattice parameters, scale factors, overall isotropic displacement parameters, Lorentzian isotropic crystallite size, Lorentzian isotropic strain, and preferred orientation. In order to determine the lattice
parameters  as a function of doping with high absolute precision,
another set of diffraction patterns was recorded in Bragg-Brentano
reflection geometry after mixing the samples with Si-powder for
calibration (Cu $K_{\alpha 1}$ radiation). Temperature dependent
measurements of the lattice parameters were undertaken in
Bragg-Brentano geometry using a He evaporation cryostate.

\subsection{Physical properties}
The electrical resistivity of the samples was measured using a standard four-point method in the temperature range 5-300~K. The electrical contacts to the cuboid-shaped specimens were made with silver paint (DuPont 4929). Magnetization measurements have been performed in the temperature
range 2-350\,K in a static magnetic field of 1\,T using a SQUID magnetometer MPMS-XL5 from Quantum
Design. Specific heat measurements in zero magnetic field have been done at 2-250\,K by means
of a relaxation technique in a Quantum Design PPMS system. Thermal conductivity and Seebeck coefficient were measured in a home-made device using a steady-state method with an SMD resistor as a heater and a Au-Chromel differential thermocouple for determining temperature gradient \cite{Hess2003}.

\section{Results and discussion} 

\subsection{Sample preparation and analysis}
\begin{figure}
\includegraphics [width=\columnwidth,clip] {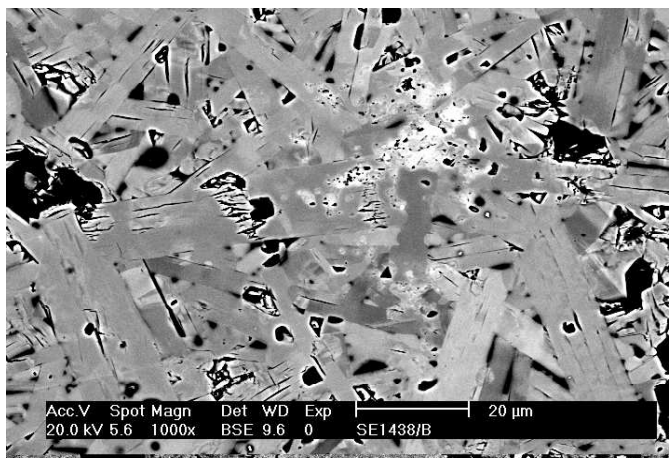}
\caption{Electron microscope (BSE mode) image of a $\rm LaO_{0.9}F_{0.1}FeAs$ sample, the grain size is some tens of micrometer; the impurity phase $\rm LaO_xF_y$ (white spots) is also visible.}
\label{fig:SEM}
\end{figure}

The prepared pellets are to a large extent porous. They exhibit grains of $\rm LaO_{1-x}F_xFeAs$ with typical dimensions of some tens of micrometers as is revealed by electron microscopy in backscattered electron imaging (BSE mode), cf. Figure~\ref{fig:SEM} for a representative example. 
We find typical mass densities which amount to about 65\% of the theoretical value.
The chemical composition of the main phase was confirmed by wavelength dispersive X-ray analysis mode. In the main phase the fluorine content was found to noticeably fluctuate between different grains with the median being close to the nominal compositions. Note that for $x\leq 0.04$ the actual F-contents are somewhat smaller than the nominal doping levels. On pellets, where the fluorine distribution was too wide, no physical measurements were carried out. In all samples second phases of different amounts were found, being largest in the high fluorine doped samples. Typical phases detected in the BSE contrast and measured in WDX mode are FeAs and $\rm LaO_xF_y$ with fluorine content higher than the oxygen one.

\begin{figure}
\includegraphics [width=\columnwidth,clip] {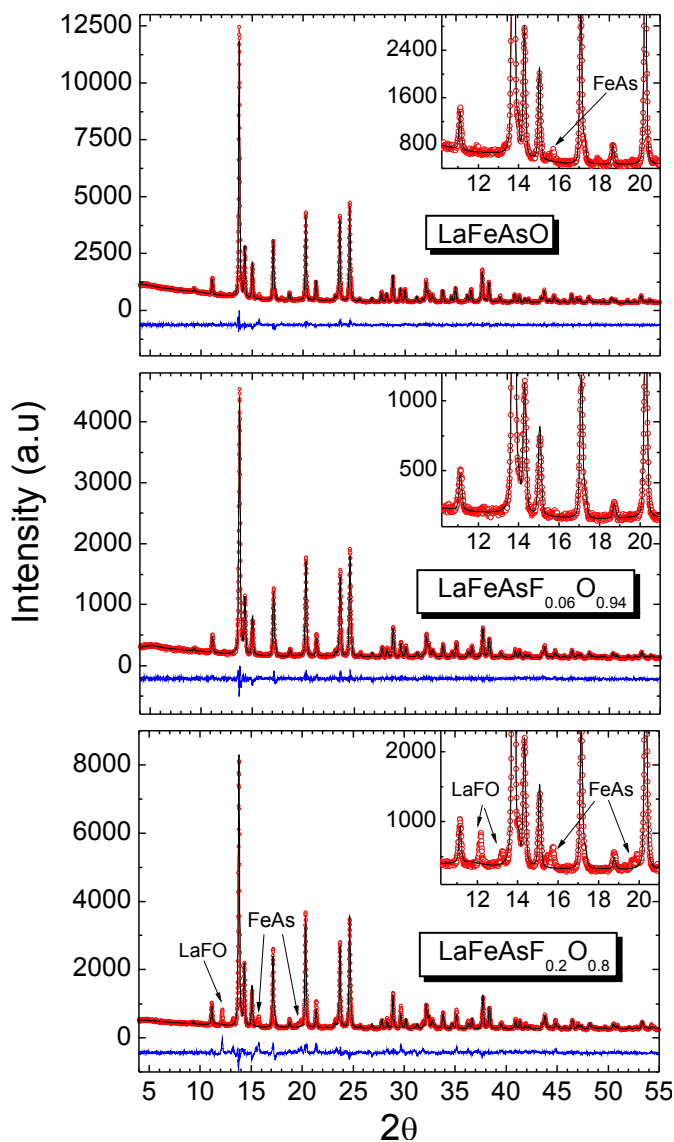}
\caption{Refined powder x-ray diffractograms for the samples with nominal composition x=0, 0.06 and 0.2. The red line corresponds to the calculated profile and the blue one to the difference between the calculated and observed profiles. The remaining peaks in the difference line for the compound with x=0.2 refers to impurity phases. The inset shows in detail the region where the impurity phases peaks are present.}
\label{fig:rietveld}
\end{figure}
Representative results of the Rietveld refinement are shown in figure \ref{fig:rietveld}. The majority of the diffraction peaks are explained by the phase $\rm LaFeAsO$ which has a tetragonal structure with space group $P 4/n m m$, in agreement with previous reported data~\cite{Sefat2008}. The remaining peaks correspond to impurity phases $\rm FeAs$ and $\rm LaFO$. The quality of the overall refinement, i.e. $\chi^2$, of all samples was found between 0.947 and 2.68 (corresponding to the samples with $x=0.125$ and $x=0.2$, respectively) which indicates a satisfactory fit.
As can be seen in Figure~\ref{fig:rietveld}, the diffraction peaks of both impurity phases are well resolvable at the doping level $x=0.2$. These secondary phases appear with fractional concentrations lower than  about 3\% except for the samples with $x=0.04$ and $x=0.2$, where the total amount of secondary phases is found to be around 5\% and 11\%, respectively. 

\begin{figure}
\includegraphics[width=\columnwidth,clip]{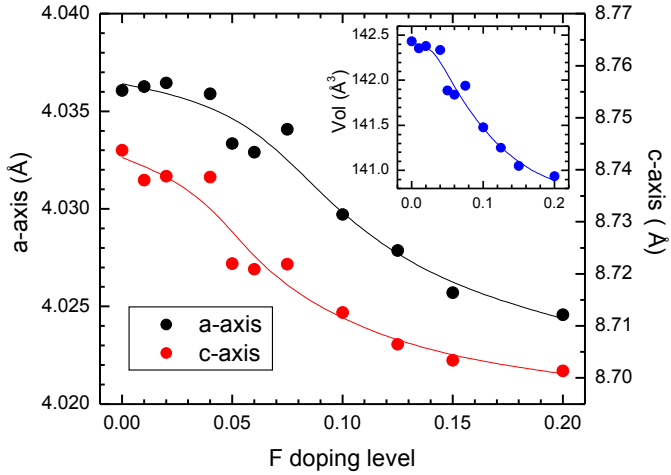}
\caption{Lattice parameters of $\rm LaO_{1-x}F_xFeAs$ as a function of F content $x$. The inset shows the doping evolution of the unit cell volume. Solid lines represent guides to the eye.}
\label{fig:latticepara}
\end{figure}

Figure \ref{fig:latticepara} shows the variation of the lattice parameters as function of the F content. It shows that both $a$ and $c$ decrease systematically by substituting $\rm O^{-2}$ by $\rm F^-$ up to the highest doping level. It appears noteworthy that
the reduction of the lattice volume seems to set in above 4\%
doping, i.e. above the critical doping level where the two low-temperature
transitions are suppressed (see the inset of Figure~\ref{fig:latticepara}).

\subsection{Physical Properties}

\begin{figure*}
\includegraphics[width=2\columnwidth,clip]{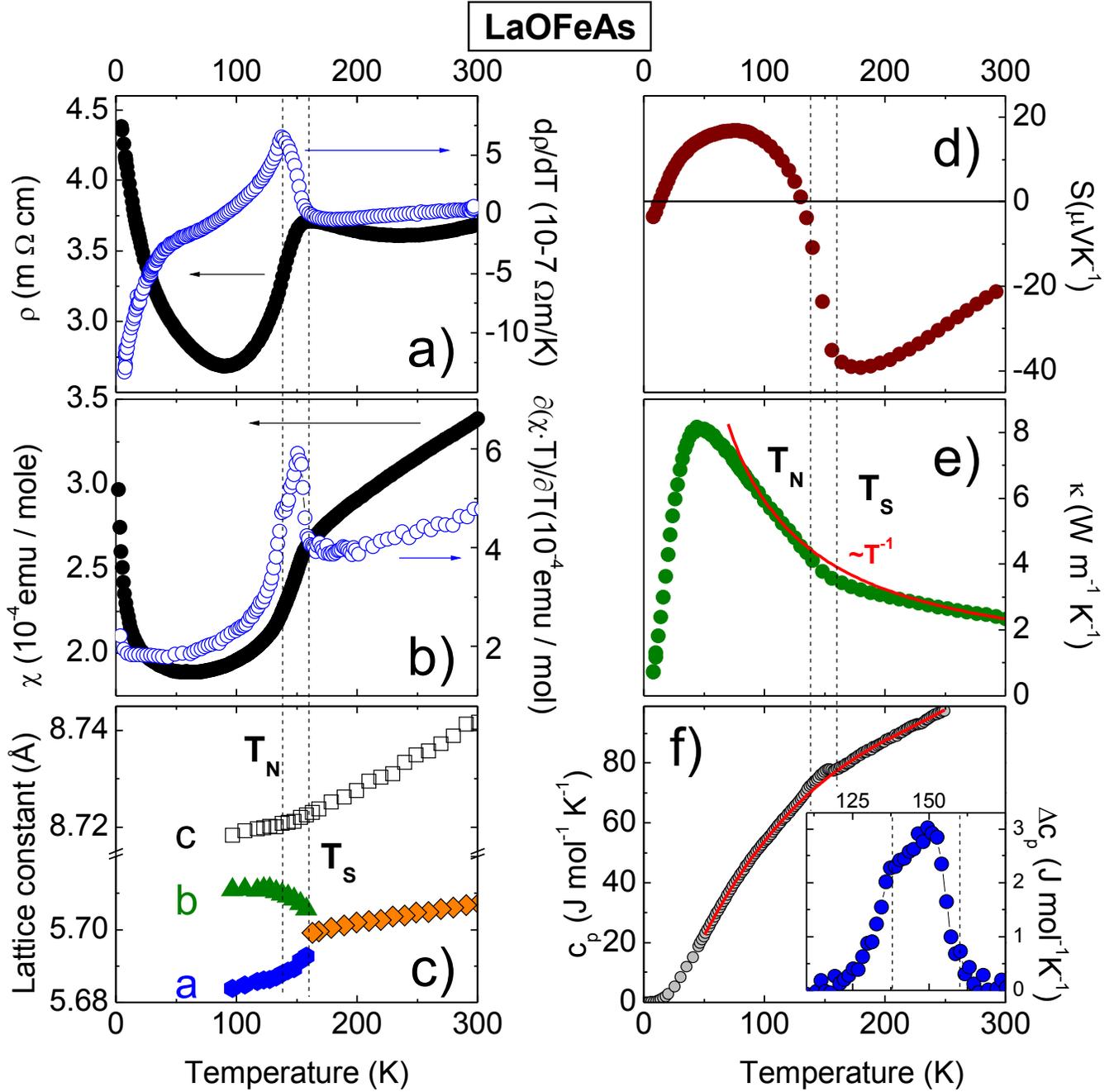}
\caption{The temperature dependence of $\rho$ (a), $\chi=M/B$ (b), lattice constants (c), $S$ (d), $\kappa$ (e)
and $c_p$ (f) for the undoped compound $\rm LaOFeAs$. Figure (a) and (b) also show the derivative $d\rho
/dT$ and the magnetic specific heat $d(\chi T)/dT$, respectively. The inset of (f) displays the
anomalous contributions $\Delta c_p$ to the specific heat as described in the text. The static
susceptibility has been measured in an external magnetic field of $B=1$\,T. The dashed lines indicate the temperatures of structural ($T_S\approx160$~K) and magnetic ($T_N\approx138$~K) transitions. The solid lines indicate indicate an usually expected $T^{-1}$-behavior of the phononic thermal conductivity (e) and the estimated phononic specific heat (f), respectively.} 
\label{fig:x0_all}
\end{figure*}

Our data of the temperature dependence of $\rho$, $\chi$, $a$- and $c$-lattice constants, $c_p$, $S$ and $\kappa$ at zero doping level $x=0$ are summarized in Figure~\ref{fig:x0_all}. The data presented in this figure show a strong impact of the structural (cf. Figure~\ref{fig:x0_all}c) and magnetic transitions at $T_S\approx160$~K and $T_N=138$~K on \textit{all} quantities shown in the figure. As has already been pointed out in References~\cite{Klauss2008,Klingeler2008,Hess2008} an intimate coupling exists between the two transitions and the electronic and magnetic properties. The structural transition is well resolved through the splitting of the (220)$_T$ reflection (tetragonal notation) into the (400)$_O$ and (040)$_O$ (orthorhombic notation) at $T_S\approx160$~K. The resistivity (see Figure \ref{fig:x0_all}a) exhibits a maximum at $T_S$ where it starts to decrease upon lowering the temperature. 
Interestingly, at the onset of long range antiferromagnetic SDW order at $T_N=138$~K \cite{Klauss2008,Cruz2008} the drop of the resistivity becomes weaker (visible through an inflection point in $\rho$ and hence a peak in $d\rho/dT$). A further decrease of temperature leads to a minimum of $\rho(T)$ at $\sim90$~K followed by a strong low-temperature upturn which is indicative of carrier localization, presumably arising from a SDW gap. The profound change of the electronic properties caused by these two transitions are also reflected in the thermopower data (Figure~\ref{fig:x0_all}d), which show a sign change near $T_N$ signalling a significant change in the Fermi surface topology and/or in the carrier relaxation time caused by the SDW order. Note, however, that the large positive contribution to $S$ which leads to the sign change apparently sets in already at $T>T_S$.

Concomitant with the resistivity drop, the static susceptibility $\chi = M/B$
(Figure~\ref{fig:x0_all}b) drops at $T_S$, indicating an enhancement of antiferromagnetic
correlations at the structural phase transition. At $T_N$ the anomaly of $\chi$ is similar but
much weaker than that at $T_S$. Both features are clearly visible if the magnetic specific heat is
considered which is proportional to $d(\chi T)/dT$ (right ordinate of Figure~\ref{fig:x0_all}b).
The data exhibit a clear jump in $d(\chi T)/dT$ at \ts . As already visible from the static
susceptibility data, the jump of the magnetic specific heat is much smaller at \tn . Both phase
transitions are also visible in the specific heat data in (Figure~\ref{fig:x0_all}f) which
measures the total entropy changes. In order to highlight these anomalies we have subtracted a
smooth background from the data (cf.~\cite{Klingeler2002}). This background was determined by
fitting the specific heat well below $T_{\rm N}$ and well above $T_{\rm S}$ by a polynomial
function as shown by the red line in Figure~\ref{fig:x0_all}f. The background specific heat mainly
reflects the phonon contribution, though an unambiguous separation of the different contributions
to $c_p$ is impossible. Using different temperature ranges for the determination of the background
and/or choosing different fit functions does not change the results significantly. The obtained
anomalous contributions to the specific heat $\Delta c_p$ are shown in the inset of
Figure~\ref{fig:x0_all}f. The data confirm a jump in $c_p$ at \ts\ which is indicative of a second
order phase transition and an additional anomaly at \tn . The similarity between $d(\chi T)/dT$
and the measured specific heat anomaly indicates that the total
entropy change connected with both transitions is proportional to that of magnetic origin. Note,
however, that the former comprises contributions from structural and charge degrees of freedom in
addition to the magnetic ones.

The strong impact of the transitions on the lattice properties is reflected in the thermal conductivity $\kappa$, which is presented in Figure~\ref{fig:x0_all}e. According to the Wiedemann-Franz law applied on the resistivity data, the electronic contribution to $\kappa$ is negligible and hence $\kappa$ of {\lao} is of phononic origin. The most prominent feature of the measured data is a significant deviation from a $\sim T^{-1}$ decrease which is usually expected for phononic heat conductivity at higher temperature. Instead, a strong suppression connected with a kink-like anomaly near $T_N$ and $T_S$ is observed which signals the sudden freezing of fluctuations connected with the transitions \cite{Hess1999,Hess2003}.

\begin{figure}
\includegraphics[width=1\columnwidth,clip]{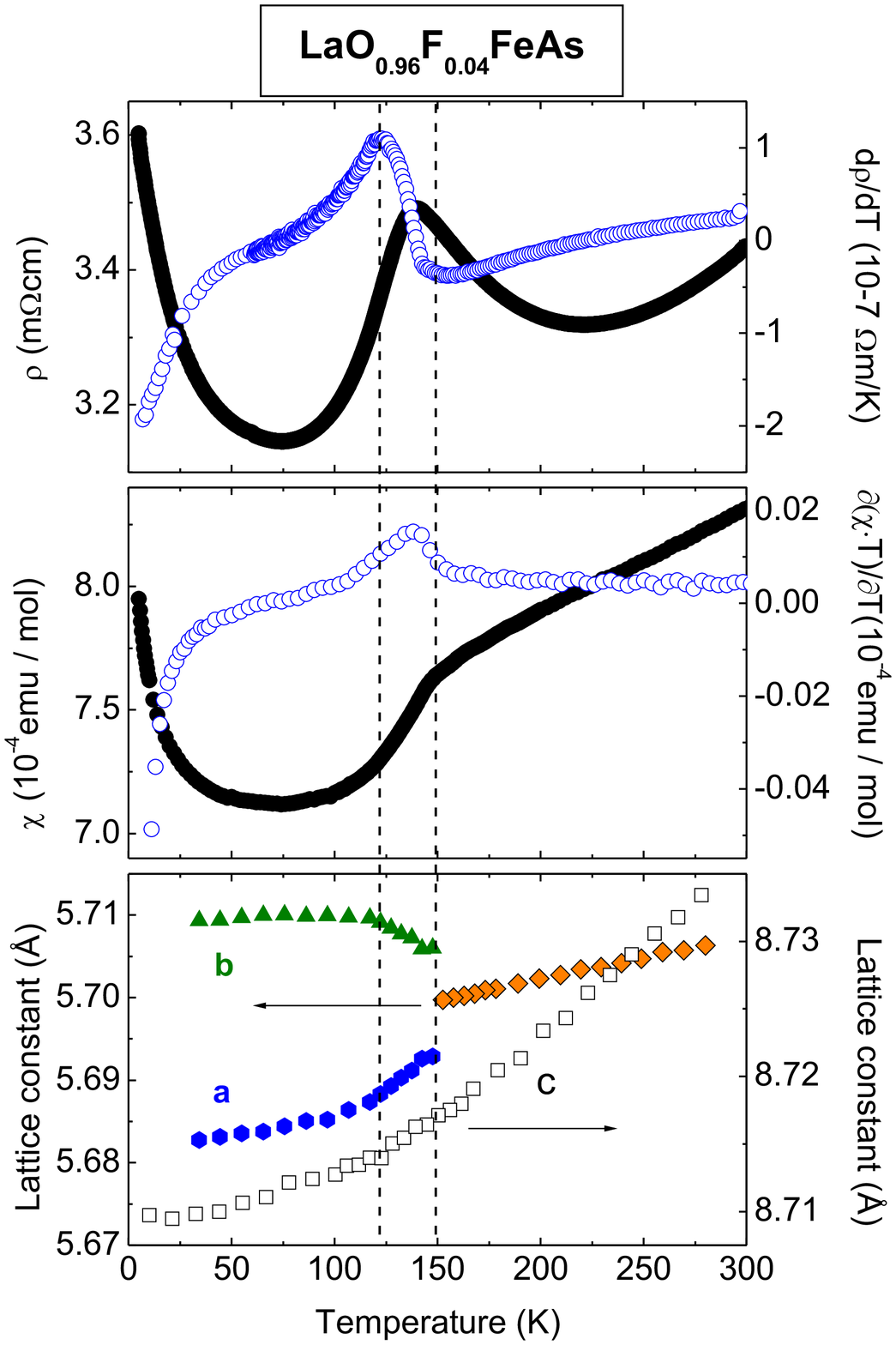}
\caption{Temperature dependence of $\rho$, $\chi=M/B$, $d\rho
/dT$, the magnetic specific heat $d(\chi T)/dT$ and the lattice constants for the highest
non-superconducting doping level $x=0.04$. }
\label{fig:x4_all}
\end{figure}

Figure~\ref{fig:x4_all} summarizes our data for $\rho$, $\chi$, the lattice constants as
well as $d\rho /dT$ and the magnetic specific heat $d(\chi T)/dT$ for the highest non-superconducting doping level $x=0.04$. A first glance on the figure suggests that doping $\rm LaO_{1-x}F_xFeAs$ up to a level of $x\approx 0.04$ leaves the qualitative properties of the material essentially unchanged. The structural transition is still clearly present as can be inferred from the orthorhombic splitting with a similar magnitude as in the undoped case. As in the undoped case, the structural transition leads to clear anomalies in $\rho$ and $\chi$. Note, however, that the transition occurs at a somewhat lower temperature $T_S\approx150$~K and that the connected anomalies in $\rho$ and $\chi$ are less pronounced and broadened. The resistivity maximum even occurs a few Kelvin below $T_S$. A similar conclusion holds for the onset of long range SDW order, which sets in at $T_N\approx 122$~K \cite{Luetkens2009a}. While $T_N$ is still well reflected through the inflection point of $\rho$, a corresponding anomaly is barely resolveable in the magnetic susceptibility.

\begin{figure}
\includegraphics[width=1\columnwidth,clip]{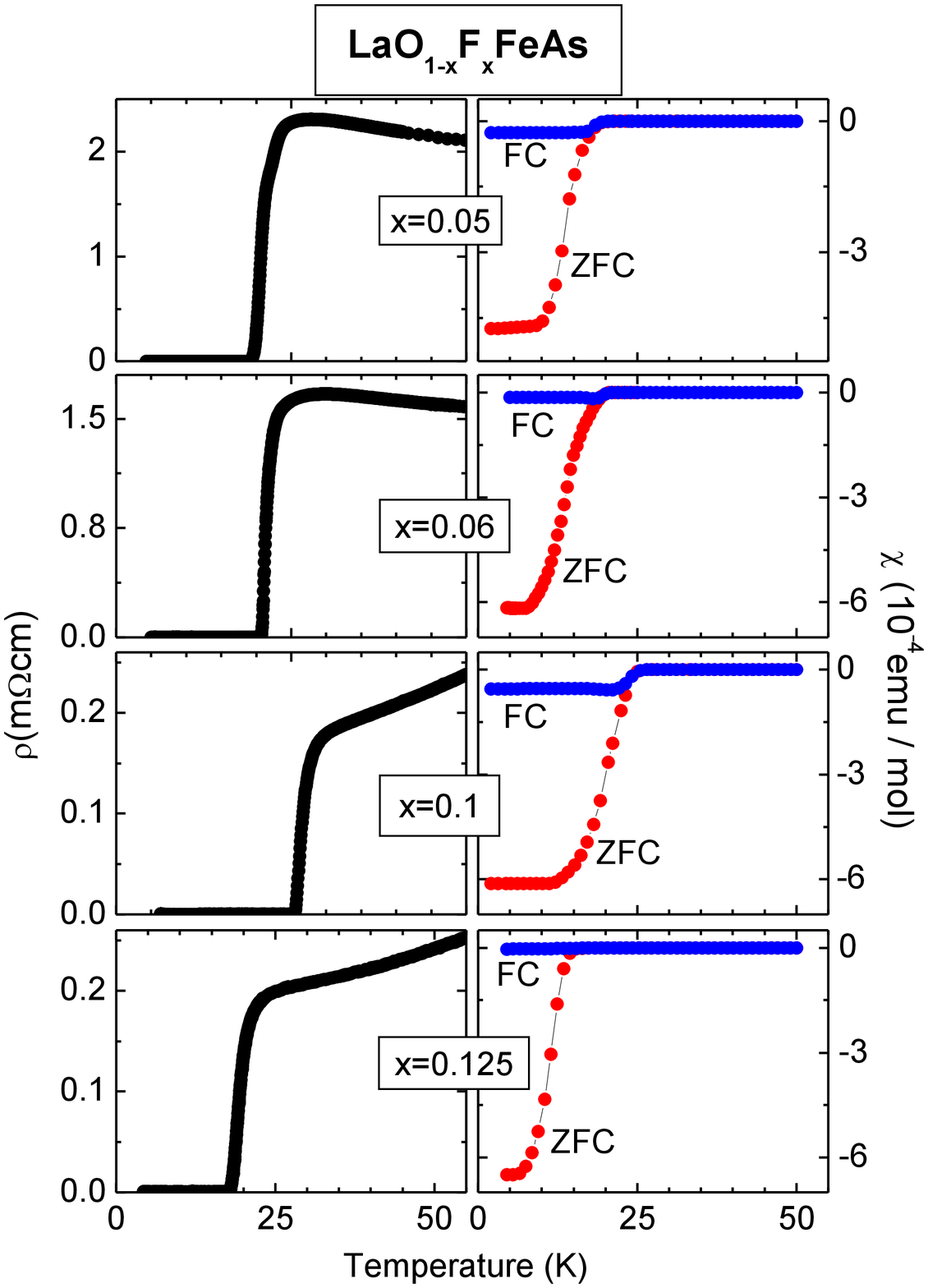}
\caption{Temperature dependence of $\rho$, $\chi$ in the vicinity of $T_c$ for representative
doping levels in the superconducting regime of the phase diagram. Magnetic measurements have
been performed in a field of 2\,mT.}
\label{fig:tcs}
\end{figure}

Superconductivity occurs abruptly with a rather high $T_c=20.6$~K at a slightly higher doping level $x=0.05$, and the superconducting phase persists up to the highest studied doping level $x=0.2$. Thereby $T_c$ increases with doping up to 26.8~K at $x=0.1$ and then quickly diminishes with further increasing $x$ down to $T_c=19.4$~K at $x=0.125$ and $T_c\approx10$~K for $x=0.15$ and 0.2. 
All these superconducting samples show a strong diamagnetic response in
zero field cooled (ZFC) as well as in field cooled (FC) measurements (cf. Figure~\ref{fig:tcs}). Note, that for $x=0.15$ and $x=0.2$, the diamagnetic signal becomes significantly weaker.

In the following we describe the magnetic, structural and electrical transport properties for
two selected and representative concentrations.
\begin{figure}
\includegraphics[width=1\columnwidth,clip]{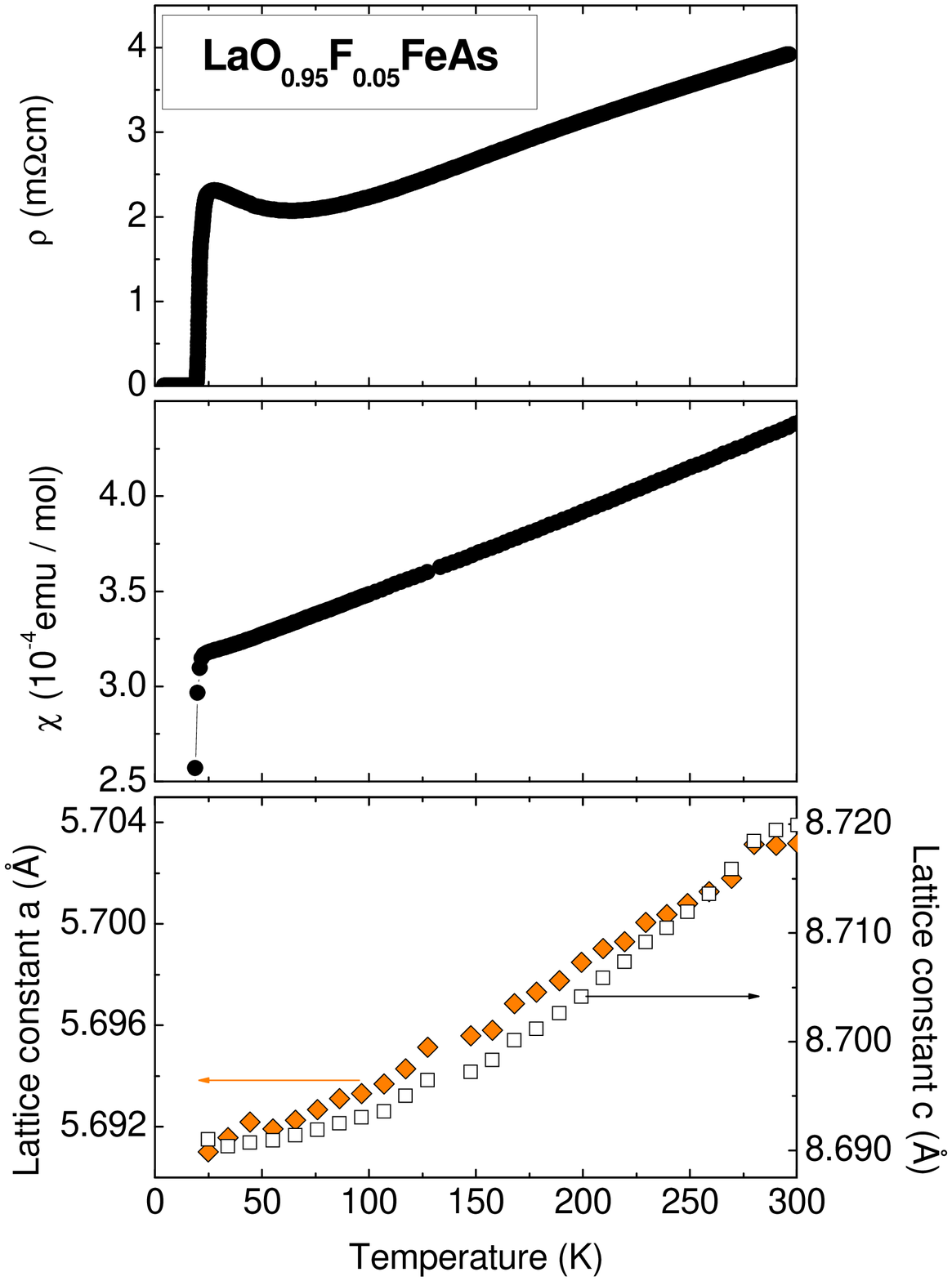}
\caption{Temperature dependence of $\rho$, $\chi=M/B$, and the lattice constants for the
lowest superconducting doping level $x=0.05$. The static susceptibility shown has been obtained in
an external field of 1\,T (cf. Figure~\ref{fig:tcs} for low field data). No structural phase transition is observed. Note, that the orthorhombic notation is used for lattice constant $a$.}
\label{fig:x5_all}
\end{figure}
Figure~\ref{fig:x5_all} summarizes our results for the lattice parameters, $\rho$ and $\chi$ at $x=0.05$, i.e. at the lowest doping level where superconductivity occurs. The structural data clearly show that the transition to the low-temperature orthorhombic phase is absent in the investigated temperature range. Note, that also static or slowly fluctuating magnetism is clearly absent as revealed by $\mu$SR experiments \cite{Luetkens2009a}. Interestingly, the suppression of the structural/magnetic transitions and the occurrence of superconductivity is accompanied by strong changes of the normal state resistivity.
A low-temperature upturn ($T\lesssim 60$~K) is still present just above $T_c=20.6$~K, which is reminiscent of the low-$T$ upturn of the non-superconducting low-doping compounds. At higher temperature, however, the clear features at $\sim150$~K of the non superconducting samples have vanished. The absence of the transition is also evident in the magnetic susceptibility: in the normal state $\chi(T)$ increases with rising temperature linearly up to room temperature. This increase is highly unusual and indicates antiferromagnetic or singlet correlations with a large energy scale \cite{Klingeler2008,Berciu2008}. An increase of the F-content up to $x=0.075$ does not lead to significant changes except a slight reduction of the low-$T$ upturn \cite{Hess2008} in the resistivity and the occurrence of a small positive curvature in the susceptibility \cite{Klingeler2008}.

\begin{figure}
\includegraphics[width=1\columnwidth,clip]{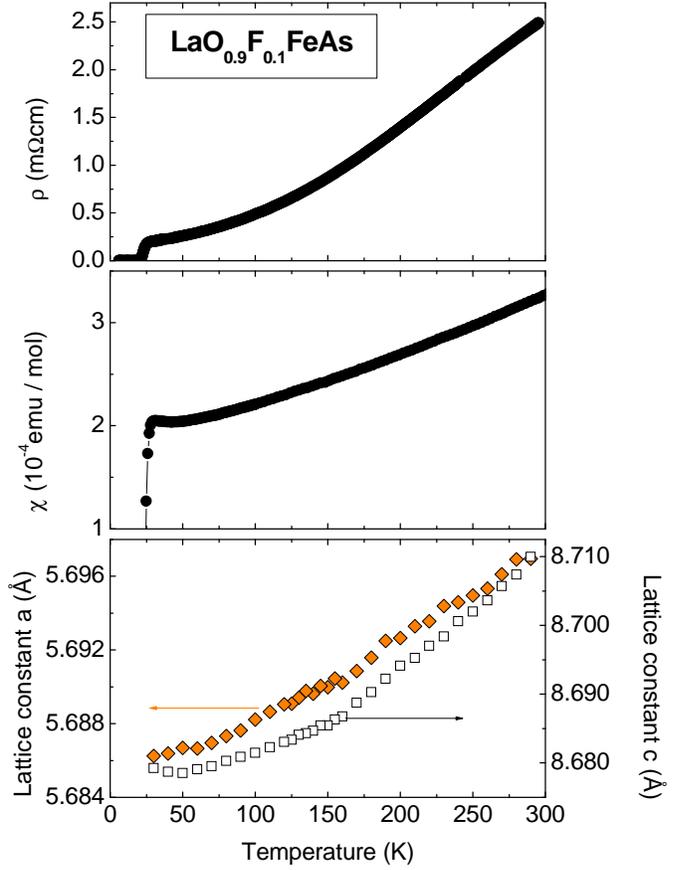}
\caption{Temperature dependence of $\rho$, $\chi=M/B$, and the lattice constants for the
doping level $x=0.10$ with highes $T_c$. The static susceptibility shown has been obtained in an
external field of 1\,T (cf. Figure~\ref{fig:tcs} for low field data).
Temperature dependence of $\rho$, $\chi=M/B$, and the lattice constants for the superconducting sample with the intermediate doping level $x=0.10$. The static susceptibility shown has been obtained in an
external field of 1\,T (cf. Figure~\ref{fig:tcs} for low field data). The sample possesses the highest $T_c=26.8$~K of all series. No structural transition is observed. Note, that the orthorhombic notation is used for lattice constant $a$.}
\label{fig:x10_all}
\end{figure}

The highest critical temperature of our $\rm LaO_{1-x}F_xFeAs$ samples occurs at the doping level $x=0.1$. Also for this compound no indication of a structural transition towards the low-temperature orthorhombic phase is observed (cf. the structural data in Figure~\ref{fig:x10_all}). The magnetic susceptibility is qualitatively and quantitatively very similar to that of the lower superconducting doping levels. The only change is a slightly more pronounced positive curvature. A drastic systematic change is present, however, in the electrical resistivity $\rho(T)$ at doping levels $0.1\leq x\leq0.2$. Here, instead of the low-$T$ upturn, we observe $\rho(T)=\rho_0+AT^2$ ($\rho_0=\mathrm{const}$) from just above $T_c$ up to $\sim200$~K \cite{Hess2008}, consistent with previous data \cite{Sefat2008} for $x=0.11$. 

\subsection{Discussion}
The physical properties exhibit a systematic change as a function of doping. Apparently, there exist three major regions in the electronic phase diagram of $\rm LaO_{1-x}F_xFeAs$ as a function of doping level $x$. At low doping levels $0\leq x\leq 0.04$ the material is a non superconducting metal with long range SDW order as the ground state. As is revealed by our data, the transition towards the SDW state and the accompanying structural transition has a strong impact on the physical properties of the system. More precisely, our data show that the transitions have a combined effect on the charge degrees of freedom (resistivity, thermopower), magnetism (magnetic susceptibility) and structure (diffraction, thermal transport). At slightly higher doping levels ($0.05\leq x\leq0.075$) the transitions are suppressed in favour of superconductivity at rather high temperature $T_c>20$~K. This clearly suggests that SDW order and superconductivity compete in the system. In this second region of the phase diagram, no traces of static magnetism is observed in $\mu$SR experiments \cite{Luetkens2009a}, in contrast to the Sm-based pendant of the system, $\rm SmO_{1-x}F_xFeAs$, where SDW order and superconductivity apparently may coexist \cite{Drew2009,Drew2008}. Despite this interesting finding, the resistivity shows a low temperature upturn just before the onset of superconductivity, which is reminiscent of the low-temperature upturn of the non superconducting species. The third region of the phase diagram comprises even higher doping levels $0.1\leq x\leq 0.2$, which includes the finding of maximum $T_c=26.8$~K at $x=0.1$ and also the lowest $T_c\approx 10$~K at $x=0.15$ and $x=0.2$. Here, instead of a the low-temperature upturn we observe a monotonically increasing resistivity curve with $\rho(T)=\rho_0+AT^2$ ($\rho_0=\mathrm{const}$) at low temperature.

\section{Summary}
We have prepared polycrystals of the new superconducting material $\rm LaO_{1-x}F_xFeAs$. The characterization of the samples clearly shows that the materials are of high quality in the sense of phase purity and systematic variation of the lattice constants as a function of the fluorine doping level. Our investigation of the structural, thermodynamic and transport properties of the samples reveal a non superconducting phase at $0\leq x\lesssim0.04$ which for $x=0$ is characterized by a structural transition towards an orthorhombic distortion at $T_s\approx 160$~K  and antiferromagnetic spin order at $T_N\approx138$K. Both transitions have a strong impact on electronic, magnetic and structural degrees of freedom. Remarkably, these transition temperatures are only weakly doping dependent up $x\lesssim 0.04$. At slightly higher doping level ($x\geq0.05$) superconductivity emerges with a concurrent suppression of the magnetic and structural transitions. Upon increasing the F-doping a maximum $T_c=26.8$~K is observed at $x=0.1$ followed by a decrease with $T_c\approx10$~K at $x\geq 0.15$. While the electrical resistivity shows pronounced doping dependence in the normal state, the magnetic susceptibility is only weakly doping dependent.

\section{Acknowledgments}
M. Deutschmann, S. M\"uller-Litvanyi, R. M\"uller, and S. Ga{\ss} for technical support. This work has been supported by the
Deutsche Forschungsgemeinschaft, through BE1749/12, through FOR 538 (BU887/4), and through SFB 608.

%
%

\begin{thebibliography}{32}

\bibitem{Kamihara2008}
Y.~Kamihara, T.~Watanabe, M.~Hirano, H.~Hosono, J. Am. Chem. Soc.
  \textbf{130} (2008), 3296.

\bibitem{Chen2008a}
X.H. Chen, T.~Wu, G.~Wu, R.H. Liu, H.~Chen, D.F. Fang, Nature
  \textbf{453} (2008), 761.

\bibitem{Cheng2008}
P.~Cheng, L.~Fang, H.~Yang, X.~Zhu, G.~Mu, H.~Luo, Z.~Wang, H.H. Wen, Science
  in China G \textbf{51} (2008), 719.

\bibitem{Liu2008a}
R.H. Liu, G.~Wu, T.~Wu, D.F. Fang, H.~Chen, S.Y. Li, K.~Liu, Y.L. Xie, X.F.
  Wang, R.L. Yang et~al., Phys. Rev. Lett. \textbf{101} (2008), 087001.

\bibitem{Ren2008b}
Z.A. Ren, J.~Yang, W.~Lu, W.~Yi, G.C. Che, X.L. Dong, L.L. Sun, Z.X. Zhao, Materials Research Innovations \textbf{12} (2008), 105.

\bibitem{Ren2008c}
Z.A. Ren, W.~Lu, J.~Yang, W.~Yi, X.L. Shen, Z.C. Li, G.C. Che, X.L. Dong, L.L.
  Sun, F.~Zhou et~al., Chin. Phys. Lett. \textbf{25} (2008), 2215.

\bibitem{Boeri2008}
L.~Boeri, O.V. Dolgov, A.A. Golubov, Phys. Rev. Lett. \textbf{101} (2008), 026403.

\bibitem{Haule2008a}
K.~Haule, J.H. Shim, G.~Kotliar, Phys. Rev. Lett. \textbf{100} (2008), 226402.

\bibitem{Mazin2008}
I.I. Mazin, D.J. Singh, M.D. Johannes, M.H. Du, Phys. Rev. Lett. \textbf{101} (2008).

\bibitem{Drechsler2008}
S.L. Drechsler, M.~Grobosch, K.~Koepernik, G.~Behr, A.~K\"{o}hler, J.~Werner,
  A.~Kondrat, N.~Leps, C.~Hess, R.~Klingeler et~al., Phys. Rev. Lett. \textbf{101} (2008), 257004.

\bibitem{Grafe2008}
H.J. Grafe, D.~Paar, G.~Lang, N.J. Curro, G.~Behr, J.~Werner,
  J.~Hamann-Borrero, C.~Hess, N.~Leps, R.~Klingeler et~al., Phys. Rev. Lett.
  \textbf{101} (2008), 047003.

\bibitem{Fuchs2008a}
G. Fuchs, S.-L. Drechsler, N. Kozlova, G. Behr, A. K\"{o}hler, J. Werner, K. Nenkov, R. Klingeler, J. Hamann-Borrero, C. Hess, A. Kondrat, M. Grobosch, A. Narduzzo, M. Knupfer, J. Freudenberger, B. B\"{u}chner, and L. Schultz, Phys. Rev. Lett. \textbf{101} (2008), 237003.

\bibitem{Hunte2008}
F.~Hunte, J.~Jaroszynski, A.~Gurevich, D.C. Larbalestier, R.~Jin, A.S. Sefat,
  M.A. McGuire, B.C. Sales, D.K. Christen, D.~Mandrus, Nature \textbf{453} (2008), 903.

\bibitem{Dai2008}
X.~Dai, Z.~Fang, Y.~Zhou, F.~chun Zhang, Phys. Rev. Lett. \textbf{101} (2008), 057008.

\bibitem{Kuroki2008}
K.~Kuroki, S.~Onari, R.~Arita, H.~Usui, Y.~Tanaka, H.~Kontani, H.~Aoki, Phys. Rev. Lett. \textbf{101} (2008), 087004.

\bibitem{Han2008}
F.~Han, X.~Zhu, G.~Mu, P.~Cheng, H.H. Wen (2008), Phys. Rev. B \textbf{78} (2008), 180503(R)

\bibitem{Lee2008}
P.A. Lee, X.G. Wen, Phys. Rev. B \textbf{78}(2008), 144517.

\bibitem{Cruz2008}
C.~de~la Cruz, Q.~Huang, J.W. Lynn, J.~Li, W.R. II, J.L. Zarestky, H.A. Mook,
  G.F. Chen, J.L. Luo, N.L. Wang et~al., Nature \textbf{453} (2008), 899.

\bibitem{Klauss2008}
H.H. Klauss, H.~Luetkens, R.~Klingeler, C.~Hess, F.J. Litterst, M.~Kraken, M.M.
  Korshunov, I.~Eremin, S.L. Drechsler, R.~Khasanov et~al., Phys. Rev. Lett.
  \textbf{101} (2008), 077005. 

\bibitem{Luetkens2008}
H.~Luetkens, H.H. Klauss, R.~Khasanov, A.~Amato, R.~Klingeler, I.~Hellmann,
  N.~Leps, A.~Kondrat, C.~Hess, A.~K\"ohler et~al., Phys. Rev. Lett.
  \textbf{101} (2008), 097009.

\bibitem{Luetkens2009a}
H.~Luetkens, H.H. Klauss, M.~Kraken, F.J. Litterst, T.~Dellmann, R.~Klingeler,
  C.~Hess, R.~Khasanov, A.~Amato, C.~Baines et~al., Nature Mater. \textbf{8}, (2009),
  305.

\bibitem{Drew2009}
A.J. Drew, C.~Niedermayer, P.J. Baker, F.L. Pratt, S.J. Blundell, T.~Lancaster,
  R.H. Liu, G.~Wu, X.H. Chen, I.~Watanabe et~al., Nature Mater. \textbf{8} (2008), 310.

\bibitem{Rotter2008a}
M.~Rotter, M.~Pangerl, M.~Tegel, D.~Johrendt, Angew. Chem. \textbf{47} (2008), 7949.

\bibitem{Rotter2008b}
M.~Rotter, M.~Tegel, D.~Johrendt, Phys. Rev. Lett. \textbf{101} (2008), 107006.

\bibitem{Rotter2008}
M.~Rotter, M.~Tegel, D.~Johrendt, I.~Schellenberg, W.~Hermes, R.~Pottgen,
  Phys. Rev. B \textbf{78} (2008), 020503.

\bibitem{Drew2008}
A.J. Drew, F.L. Pratt, T.~Lancaster, S.J. Blundell, P.J. Baker, R.H. Liu,
  G.~Wu, X.H. Chen, I.~Watanabe, V.K. Malik et~al., Phys. Rev. Lett.
  \textbf{101} (2008), 097010.

\bibitem{Goko2008}
T.~Goko, A.A. Aczel, E.~Baggio-Saitovitch, S.L. Bud'ko, P.C. Canfield, J.P.
  Carlo, G.F. Chen, P.~Dai, A.C. Hamann, W.Z. Hu et~al., Preprint at
  http://arxiv.org/abs/0808.1425v1.

\bibitem{Park2008}
J.T. Park, D.S. Inosov, C.~Niedermayer, G.L. Sun, D.~Haug, N.B. Christensen,
  R.~Dinnebier, A.V. Boris, A.J. Drew, L.~Schulz et~al., Phys. Rev. Lett. \textbf{102} (2009), 117006.

\bibitem{Zhu2008a}
X.~Zhu, H.~Yang, L.~Fang, G.~Mu, H.H. Wen, Superconductor Science and
  Technology \textbf{21} (2008), 105001.

\bibitem{Souptel2007}
D.~Souptel, W.~L\"{o}ser, W.~Gruner, G.~Behr, Journal of Crystal Growth
  \textbf{307} (2007), 410.

\bibitem{Hess2003}
C.~Hess, B.~B\"{u}chner, U.~Ammerahl, A.~Revcolevschi, Phys. Rev. B \textbf{68} (2003), 184517.

\bibitem{Sefat2008}
A.S. Sefat, M.A. McGuire, B.C. Sales, R.~Jin, J.Y. Howe, D.~Mandrus, Phys. Rev.
  B \textbf{77} (2008), 174503.

\bibitem{Klingeler2008}
R.~Klingeler, N.~Leps, I.~Hellmann, A.~Popa, C.~Hess, A.~Kondrat,
  J.~Hamann-Borrero, G.~Behr, V.~Kataev, B.~B\"{u}chner, Preprint at
  http://arxiv.org/abs/0808.0708v1.

\bibitem{Hess2008}
C.~Hess, A.~Kondrat, A.~Narduzzo, J.E. Hamann-Borrero, R.~Klingeler, J.~Werner,
  G.~Behr, B.~B\"{u}chner, Preprint at
  http://arxiv.org/abs/0811.1601.

\bibitem{Klingeler2002}
R.~Klingeler, J.~Geck, R.~Gross, L.~Pinsard-Gaudart, A.~Revcolevschi,
  S.~Uhlenbruck, B.~B\"{u}chner, Phys. Rev. B \textbf{65} (2002), 174404.

\bibitem{Hess1999}
C.~Hess, B.~B\"{u}chner, M.~H\"{u}cker, R.~Gross, S.W. Cheong, Phys. Rev. B \textbf{59} (1999), R10397.

\bibitem{Berciu2008}
M.~Berciu, I.~Elfimov, G.A. Sawatzky,  Preprint at
  http://arxiv.org/abs/0811.0214.


\end{thebibliography}

\end{document}